\documentclass[12pt]{iopart}

\usepackage{graphicx}

\begin{document}

\title{Gamma-ray bursts and terrestrial planetary atmospheres}

\author{B. C. Thomas$^1$, A. L. Melott$^2$}

\address{$^1$ Department of Physics and Astronomy, Washburn University, 
  1700 SW College Ave., Topeka, KS 66621}
\ead{brian.thomas@washburn.edu}
\address{$^2$ Department of Physics and Astronomy, University of Kansas, 
1251 Wescoe Hall Dr., 1082 Malott Hall, Lawrence, KS 66045-7582}

\begin{abstract}
We describe results of modeling the effects on Earth-like planets of long-duration gamma-ray bursts (GRBs) within a few kiloparsecs.  A primary effect is generation of nitrogen oxide compounds which deplete ozone.  Ozone
 depletion leads to an increase in solar UVB radiation at the surface, enhancing DNA damage, particularly in
 marine microorganisms such as phytoplankton.  In addition, we expect increased atmospheric opacity due to 
buildup of nitrogen dioxide produced by the burst and enhanced precipitation of nitric acid.  We review here 
previous work on this subject and discuss recent developments, including further discussion of our estimates of the rates of impacting GRBs and the possible role of short-duration bursts.
\end{abstract}

\section{Introduction}
More attention has been paid in recent decades to the impact of astrophysical events on the Earth and on 
Earth-like planets which may exist elsewhere.  Impacts by asteroids and comets have been studied and connected
 with at least one major mass extinction \cite{alv80}.  Radiation events, such as large solar flares,
 supernovae, etc. have also been considered.  Gamma-ray bursts are another event in this category and were 
suggested more than a decade ago \cite{thor95} as a possible source of danger,
and later targeted \cite{smith04} as a possible accelerating mechanism for evolution by intermittent smaller 
jolts of radiation.  

Knowledge of the effects of a relatively nearby GRB on the Earth are of interest for understanding the history
 and future of life on Earth, as well as in addressing questions of where and when life may exist elsewhere in 
the Galaxy and beyond.  GRBs may have played a role in at least one mass extinction in the Earth's history 
\cite{mel04}.  Quantifying concepts such as the galactic habitable zone \cite{gonz01} require information on
 the rates and effective distances of GRBs throughout the Galaxy.

An important direct impact of a GRB on an Earth-like planet is on atmospheric chemistry due to ionization and
 dissociation of $\mathrm{O_2}$ and $\mathrm{N_2}$, leading to formation of nitrogen oxide compounds. 
 Subsequent effects include destruction of stratospheric ozone which permits solar UVB (290 -- 315 nm) to 
penetrate more effectively to the ground.  This UVB flux enhances DNA damage, especially in simple life forms
 such as phytoplankton.  In addition, atmospheric darkening may result due to greatly increased levels of 
$\mathrm{NO_2}$.  Precipitation of nitrogen oxide compounds generated by the burst as nitric acid rain may have
 complicated effects, including even a short-term enhancement of plant productivity.  Each of these effects 
has been explored in some detail by the authors and collaborators.  Full details may be found in 
\cite{mel04, thom05a,thom05b,mel05}.

We summarize here these results and discuss implications and future work.

\section{GRB Irradiation of the Earth}
No galactic GRBs have been directly observed, unless one notes that an outburst like SGR 1806-20 might appear as a short gamma-ray burst 
(which typically have harder spectra), called an SHB, if located in a nearby galaxy \cite{hurley05}. The probability of a ``near burst'' has been estimated \cite{sw02}; \cite{mel04}; \cite{dh05} with the 
conclusion that the most probable nearest long-burst GRB directed at the Earth in the last Gy
is of order 1 kpc distant.  

We estimate the likely rate and distance of GRBs directed at 
the Earth as follows \cite{mel04}.  GRB distances are large
 compared to the thickness of the disk of the Galaxy, and so we model the GRB rate as a 2D Poisson process, 
after \cite{sw02}.  Based on a review of long duration GRB rate information we normalized the $z = 0$ rate 
to $0.44~\mathrm{Gpc^{-3}y^{-1}}$ \cite{guet05}.  Scaling to the blue luminosity density (young, massive stars, as appropriate to the rapid recycling approximation)
 we estimate the {\it local} rate as about $6 \times 10 ^{-11}~\mathrm{y^{-1}kpc^{-2}}$.

Recently, there have been arguments that the precursor events of long bursts (LSB) take place in low metallicity environments, with the implication that the rate at cosmologically recent times in our galaxy should be lower \cite{ln06,sg06}.  If so, the rate of LSB might be lower than estimated here.  The situation is far from clear however.  Our galaxy appears to have a history of merger events with low-metallicity satellites whose characteristics look similar to those inferred for LSB host galaxies \cite{sg06}. LSB host galaxies also frequently appear to have disturbed morphologies indicating interaction with another galaxy \cite{wb05}.  For example, the (very low metallicity) Sagittarius dwarf galaxy is in a close orbit with period 0.85-1 Gy, and apogalactica 12-15 kpc \cite{iw97,lj05}.  Studies of stellar populations indicate evidence for a very recent merger event in bulge stars \cite{vl03} and another event is suggested in the outer disk by Cepheids (large mass, low metallicity) \cite{yc05}. Taken together, these suggest accretion events by low-mass, low-metallicity dwarf galaxies into the Milky Way possibly every few 100 My. Also, there is also good evidence for at least one very young GRB remnant in our Galaxy \cite{atoy06}.

The situation with SHB is different.  Formerly they were neglected due to low total luminosity. As we write, rate estimates are in flux due to the accumulation of new information from Swift \cite{geh04}. There is evidence that for at least one class of SHB the rate has been constant or increasing \cite{ng05,gp05,ba05}
while others are nearly local \cite{tan05}. The higher rate of SHB partially compensates for their lower luminosity. Based on these results, we estimate that the expected rate at a given local fluence for SHB is not more than one order of magnitude lower than the LSB rate we model here. Also, as we shall describe later, bursts with higher-energy photons have a greater effect for the same total burst energy level \cite{ejz06}.  This also enhances the expected effect of SHB.
The situation is fluid, and will stabilize within a year or two with increased rate information.  In the meantime we will use the rate previously adopted, discussing how to scale events at other fluences.

Modeling as a Poisson 
process (see Appendix B of \cite{sw02}), the most probable nearest long duration GRB aimed at us in the last
Gy is about 2 kpc away.  Since the recieved fluence goes as $R^{-2}$ and the number of events goes roughly 
as $R^2$, the probability of a burst goes as the inverse of the fluence.  This is, of course, only considering
 distances smaller than the scale of our galaxy.  Beyond this the estimate would be more complicated and is
 irrelevant for major impact on the biosphere.  See Table~\ref{tb1} for the probability of a GRB impacting the Earth in the
 last Gy at several distances.  On that timescale, 2 kpc is the expected {\it mean} distance to the nearest burst and there is a 55\% probability that the Earth has been hit by a (long-burst) GRB from that distance or {\it closer}.
A Gy is a relevant timescale since both the existence of a high-oxygen atmosphere (required for the presence of 
an ozone layer)
and (not coincidentally) a reasonable fossil record exist on Earth for this order of magnitude
timescale.  These estimates are independent of the beaming angle, since they are scaled directly from
observation. 
We shall present results for events of typical expected fluence and
and note how one can scale the effects.

The attenuation length for photons of a few hundred keV is of order 100m in air at Standard Temperature
and Pressure.  Consequently effects at the surface of terrestrial planets are primarily indirect, except for
 a brief
flash of moderate intensity UV at the ground. Of course direct effects may be much greater for poorly
shielded planets such as Mars.  Radiative transfer for events such as this has been explored in
some detail \cite{smith04}.  For planets like the Earth, keV-MeV photons have their greatest effect by
breaking the very strong triple bond in $\mathrm{N_2}$, which comprises about 80\% of the atmosphere.  It is 
normally quite
stable, which is why ``fixing'' nitrogen, i.e. converting it to some other, more available form, is a  major
limiting factor for biological productivity \cite{sch97}.  Ionization and dissociation of $\mathrm{N_2}$,
and also $\mathrm{O_2}$ opens a number of reaction pathways normally blocked by the stability of these molecules.
A wide range of reaction products are formed in a complicated interaction involving atmospheric
chemistry, transport, and the effects of sunlight.  This is discussed in moderate detail elsewhere
\cite{thom05b}. The most important effects for the biosphere are great changes in the concentration of
NO, $\mathrm{NO_2}$, and $\mathrm{O_3}$ (ozone). 

\cite{rei78} discussed in a general way the potential for severe
stress of the biosphere by any strong ionization event.  The three main effects are rainout of
nitric acid ($\mathrm{HNO_3}$), which helps return the atmosphere to its pre-burst state; some opacity to
visible light by $\mathrm{NO_2}$, which is a brown gas; and the action of NO and $\mathrm{NO_2}$ to catalyze
 the destruction
of $\mathrm{O_3}$.
The comparative importance of these three for
typical GRB ionization events were studied \cite{mel05} and the ozone depletion is clearly of major
importance.  The usual ozone column density in the stratosphere is not large, but it very strongly
absorbs in the UVB (290 -- 315 nm).  UVB is emitted by the Sun, and is strongly absorbed by 
proteins and by the DNA molecule, breaking bonds and altering the structure.  It is a mutagen and a carcinogen.
The existing ozone layer removes about 90\% of UVB incident at the top of the atmosphere, and has done so as
 long as the Earth has had a 
high atmospheric oxygen abundance.  The ozone concentration in the unperturbed atmosphere is less than linearly
dependent on oxygen abundance, so even lower-oxygen primordial atmospheres enjoyed considerable
protection from UVB.  \cite{mel05} noted that $\mathrm{NO_2}$ opacity might provide a perturbation toward
glaciation in a situation with that instability present, and that $\mathrm{HNO_3}$ rainout might provide
an acid stress, or paradoxically a nitrate fertilizer effect, adding a non-negligible amount
to the nitrate budget \cite{sch97}.  Because the ozone depletion effect is clearly a calculable and major
stress, we will focus on this as our diagnostic of the severity of events.

\section{Methods}
Since the primary long-term effect of a GRB on the Earth or Earth-like planet is through atmospheric chemistry,
 our studies have been conducted using an atmospheric chemistry and dynamics model.  We have focused on ozone
 depletion, subsequent increase in solar UVB flux at the ground, and DNA damage.  Our model is the Goddard Space 
Flight Center 2D atmospheric chemistry and dynamics model.  Its two dimensions are altitude and latitude.  
It is empirically based and so is most appropriate for considering effects on the Earth's atmosphere in its 
current state, but results should be roughly applicable to other times in the Earth's history, as well as to
 other planets with similar atmospheric conditions.  Full details of the model may be found in \cite{thom05b} 
and references therein.

Effects of photons from the GRB are introduced into the model via an offline calculation which computes 
atmospheric ionization profiles.  We use the Band spectrum \cite{band93} in these computations.  For results 
presented here we  take typical spectral parameter values $E_0 = 187.5~ \mathrm{keV}$, $\alpha = -0.8$, 
$\beta = -2.3$ \cite{preece00}.  Continuing work involves varying parameters such as the peak energy and
 investigating the impact on the overall results.

Increased UVB at ground level is computed using measured solar irradiance at the top of the atmosphere
 combined with ozone concentrations from the atmospheric modeling results, using the Beer-Lambert exponential
 decay law.  DNA damage is computed using UVB results convolved with a biological weighting function 
\cite{set74,smith80}.  
There is a wide variety of different weighting functions, quantifying specific effects such as the formation of pyrimidine dimers in skin cells \cite{free89}, erythema (sunburn \cite{mck87}), inhibition of phytoplankton photosynthesis \cite{cull92}, fish melanoma \cite{set93}, mortality of zooplankton and fish \cite{kou99}, and others \cite{jag85}. The spectrum one chooses is dependent upon the endpoint of interest.  We are primarily interested in the impact on simple, one-celled, transparent, marine organisms which lie at the base of the food chain.  In addition, our results for DNA damage are presented as \emph{relative} damage, a ratio of the damage rate after a burst to the normal damage rate.  This should be less sensitive to the details of the action spectrum since it measures the relative rather than absolute impact of UVB.  Finally, we limit our computations to UVB (as opposed to UVA) since ozone has little impact in other biologically relevant wavebands and we are most concerned with the heightened damage due to ozone depletion.

\section{Results}
We present here a limited set of results which give a sense of the overall effect of a GRB on the Earth. 
 Full details are presented in \cite{thom05b}.  We begin with depletion of ozone which results from catalytic
 reactions with nitrogen oxide compounds, $\mathrm{NO_y}$, most importantly $\mathrm{NO_2}$ and  $\mathrm{NO}$.
  These compounds are created when gamma-rays dissociate and ionize  $\mathrm{N_2}$ in the stratosphere, which
 reacts rapidly with  $\mathrm{O_2}$ to form NO.  Catalytic reactions then proceed to deplete  $\mathrm{O_3}$
 through the cycle:
$\mathrm{ NO + O_3 \rightarrow NO_2 + O_2 }$, $\mathrm{ NO_2 + O \rightarrow NO + O_2 }$.  The net result is:
 $\mathrm{ O_3 + O \rightarrow O_2 + O_2 }$.

\begin{figure}
\includegraphics{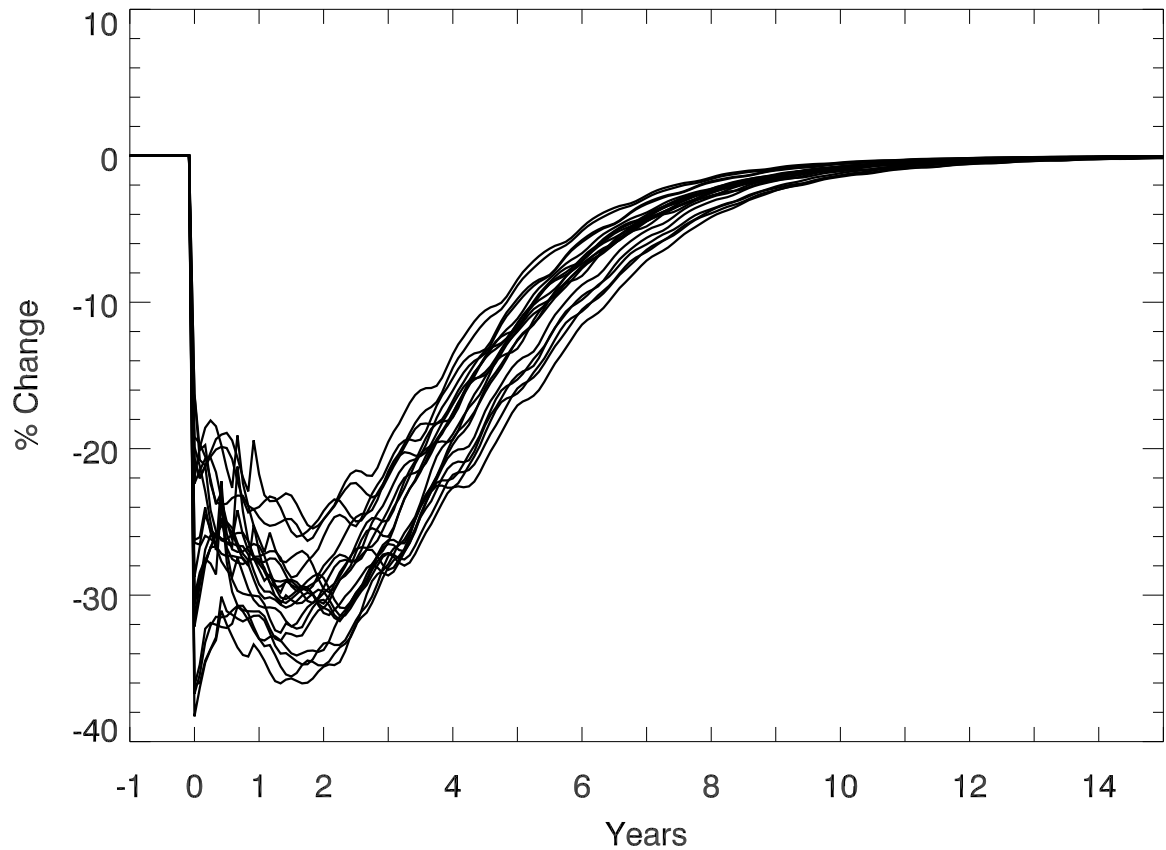} 
\caption{Percent change in globally averaged column density of $\mathrm{O_3}$, for $100~\mathrm{kJ~m^{-2}}$ bursts
 over latitudes $+90^\circ$, $+45^\circ$, the equator, $-45^\circ$, and $-90^\circ$, at the equinoxes and
 solstices. (Bursts occur at month 0.)
\label{fig:O3_perchg-glob}}
\end{figure}

In Figure~\ref{fig:O3_perchg-glob} we show globally averaged ozone depletion for a single GRB fluence of
 $100 ~\mathrm{kJ ~m^{-2}}$, occurring over 5 different latitudes ($+90^\circ$, $+45^\circ$, the equator,
 $-45^\circ$, and $-90^\circ$) and at four different times of the year (equinoxes and solstices).  This fluence 
value corresponds to a ``typical'' GRB with power $5\times10^{44}~\mathrm{W}$ (isotropic equivalent) at a 
distance of 2 kpc.  

Note that there is a wide range in maximum globally averaged depletion and the time the atmosphere takes to 
recover after the burst.  Depletions of at least 10\% are present up to 5-7 years and full recovery takes 
10-12 years.  
The maximum depletion value (38\%) occurs for a burst over the equator in September.  It is interesting to
 note that present-day globally averaged anthropogenic depletions are of the order of 3\%.

\begin{figure}
\includegraphics{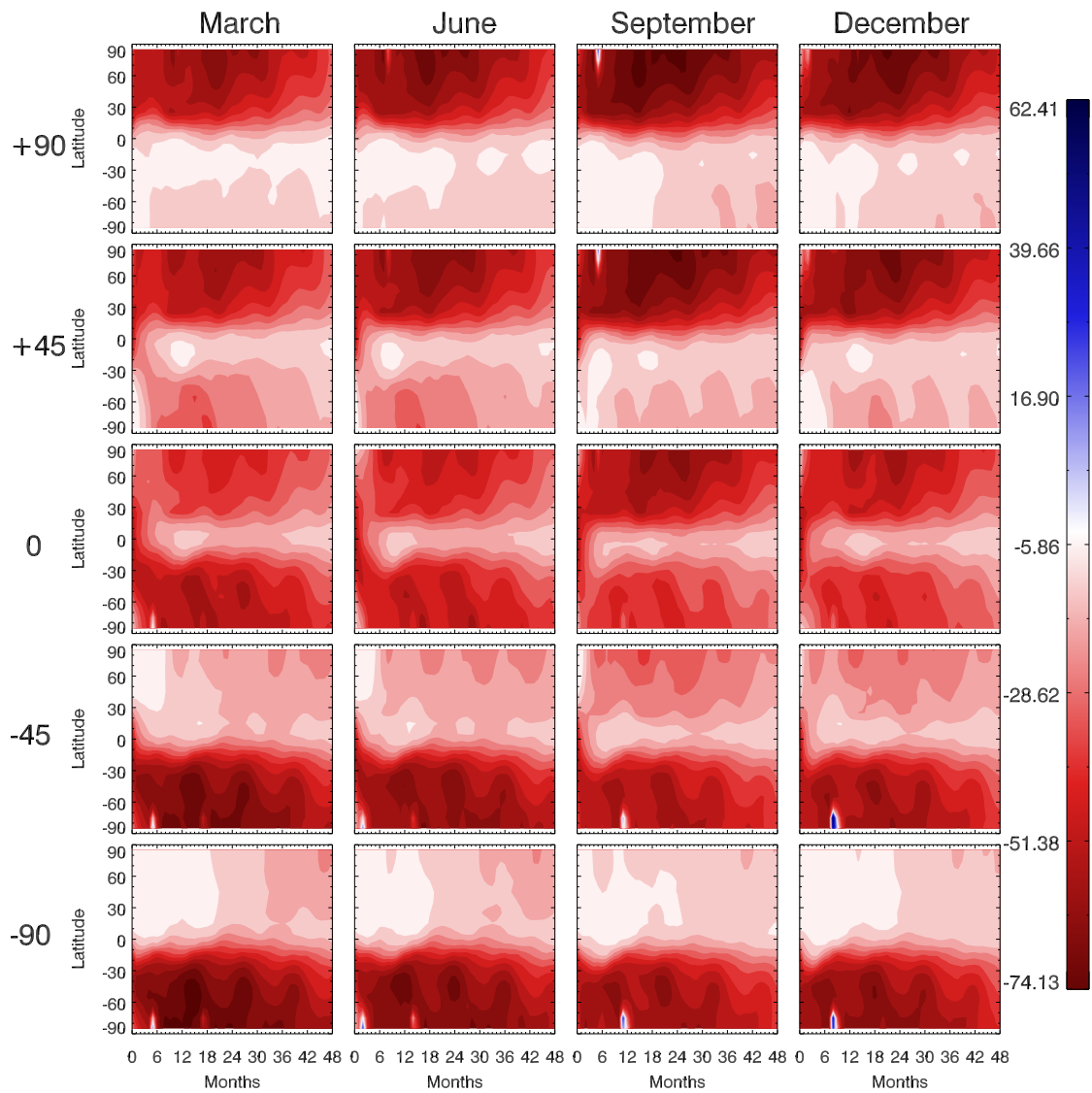}
\caption{Pointwise percent change in column density of $\mathrm{O_3}$ (comparing runs with and without burst), 
for $100~\mathrm{kJ~m^{-2}}$ bursts over latitudes $+90^\circ$, $+45^\circ$, the equator, $-45^\circ$, and
 $-90^\circ$, at the equnioxes and solstices. (Bursts occur at month 0.) 
\label{fig:O3dpl_grid} }
\end{figure}

The extent of depletion in latitude over time is shown in Figure~\ref{fig:O3dpl_grid}, for the same set of 
burst cases.  It is apparent that more intense local depletions occur for high latitude bursts, however the
 effects are mostly contained within the hemisphere.  Equatorial bursts yield more evenly 
distributed depletion.  Higher depletion at the poles in all cases is due primarily to colder temperatures 
in the
polar stratosphere.  As we will see, however, surface effects such as DNA damage by solar UVB are more intense 
at mid latitudes, due to the combination of the latitudinal variation of ozone depletion with duration and angle of sunlight.  Differences in depletion levels between events 
at different times of year are due primarily to the timing of polar spring relative to the burst, since the 
presence or absence of sunlight is a critical factor in the photochemistry governing these results.  It is 
interesting to note that the maximum depletion in the Antarctic ozone ``hole'' has been observed to be about
 60\%.  We see GRB-initiated depletion significantly greater than this over much larger latitude ranges than
 are currently observed (Fig.~\ref{fig:O3dpl_grid}).

\begin{figure}
\includegraphics{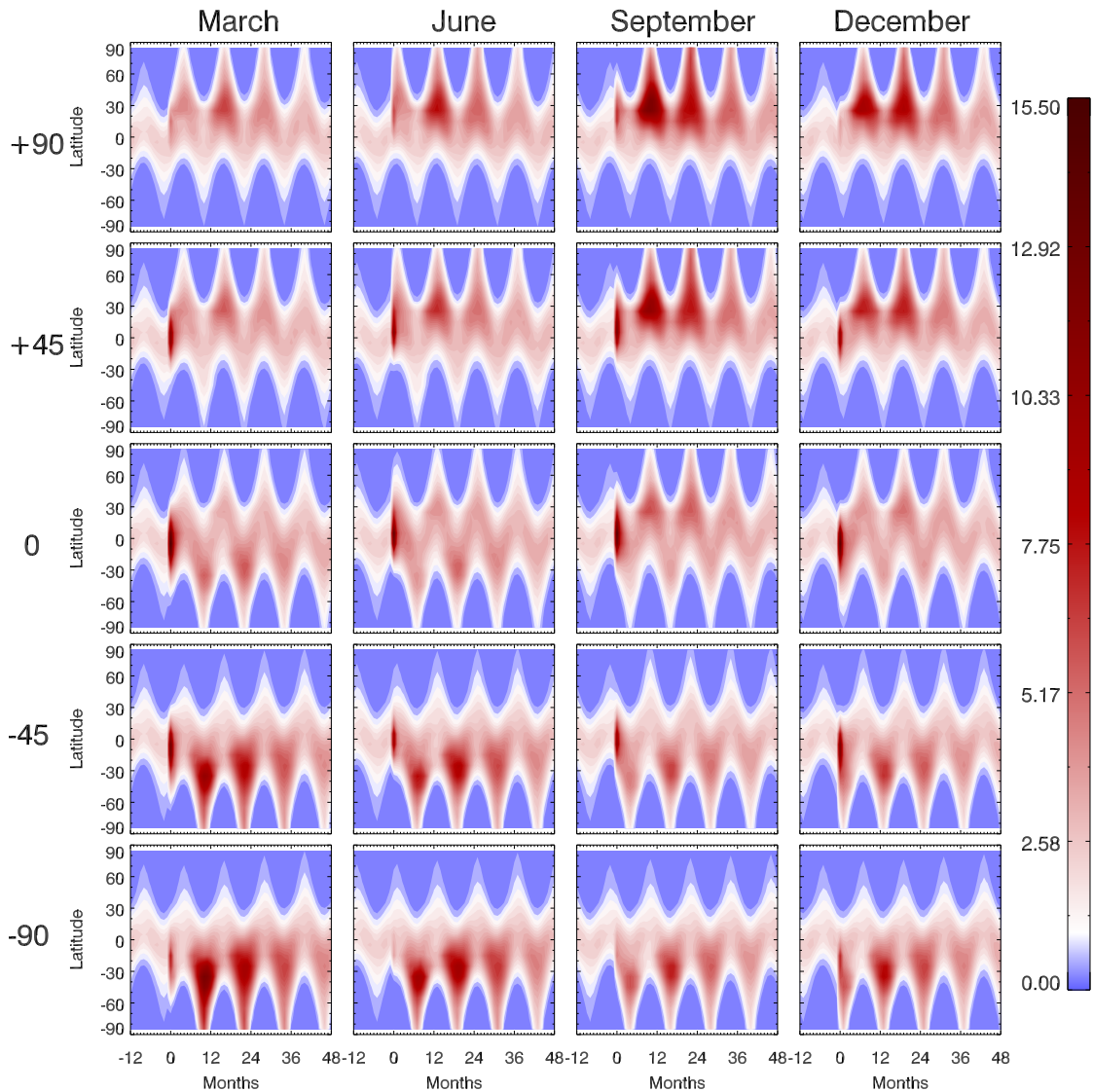}
\caption{Relative DNA damage (dimensionless), normalized by the annual global average damage in the absence of a 
GRB, for $100~\mathrm{kJ~m^{-2}}$ bursts over latitudes $+90^\circ$, $+45^\circ$, the equator, $-45^\circ$, and
 $-90^\circ$, at the equnioxes and solstices. (Bursts occur at month 0.)  Note that white is set to 1.0. 
Experiments suggest significant mortality for marine 
microorganisms should certainly exist in areas where this measure exceeds 2.
\label{fig:dna_grid} }
\end{figure}

Several effects due to the atmospheric impact of production of $\mathrm{NO_y}$ and depletion of $\mathrm{O_3}$
 have been examined.  An important effect at the Earth's surface for any land or shallow (less than 10 m) water
 dwelling life is the increase in solar UVB 
radiation which can penetrate through the reduced ozone concentrations.  We have quantified the impact of this
 increase in UVB by computing the irradiance at the surface in the presence of our reduced $\mathrm{O_3}$ levels
 and convolving this with a DNA weighting function which quantifies the effectiveness of particular wavelengths
 in damaging DNA molecules (taken from \cite{set74,smith80}).  Figure~\ref{fig:dna_grid} shows the relative DNA
 damage weighted to the annual global average before the burst for the event times and latitudes discussed 
previously.

An important point that may be noted in Fig.~\ref{fig:dna_grid} is the fact that DNA damage is greatest,
 indeed is only great for long periods of time, at mid latitudes.  This is due to a combination of the 
localized ozone depletion and the presence and angle of incident sunlight.  Polar areas have periods of 
complete darkness and even during daylight the solar incidence angle is small.  So, while the ozone depletion
 in the long-term is largest at the poles, the DNA damage in these regions is limited due to the reduced 
incident solar UVB.  It is also interesting to note that experimental results have shown that significant 
mortality for marine microorganisms, such as phytoplankton, is likely above a value of 2.0 in our units.

\begin{figure}
\includegraphics{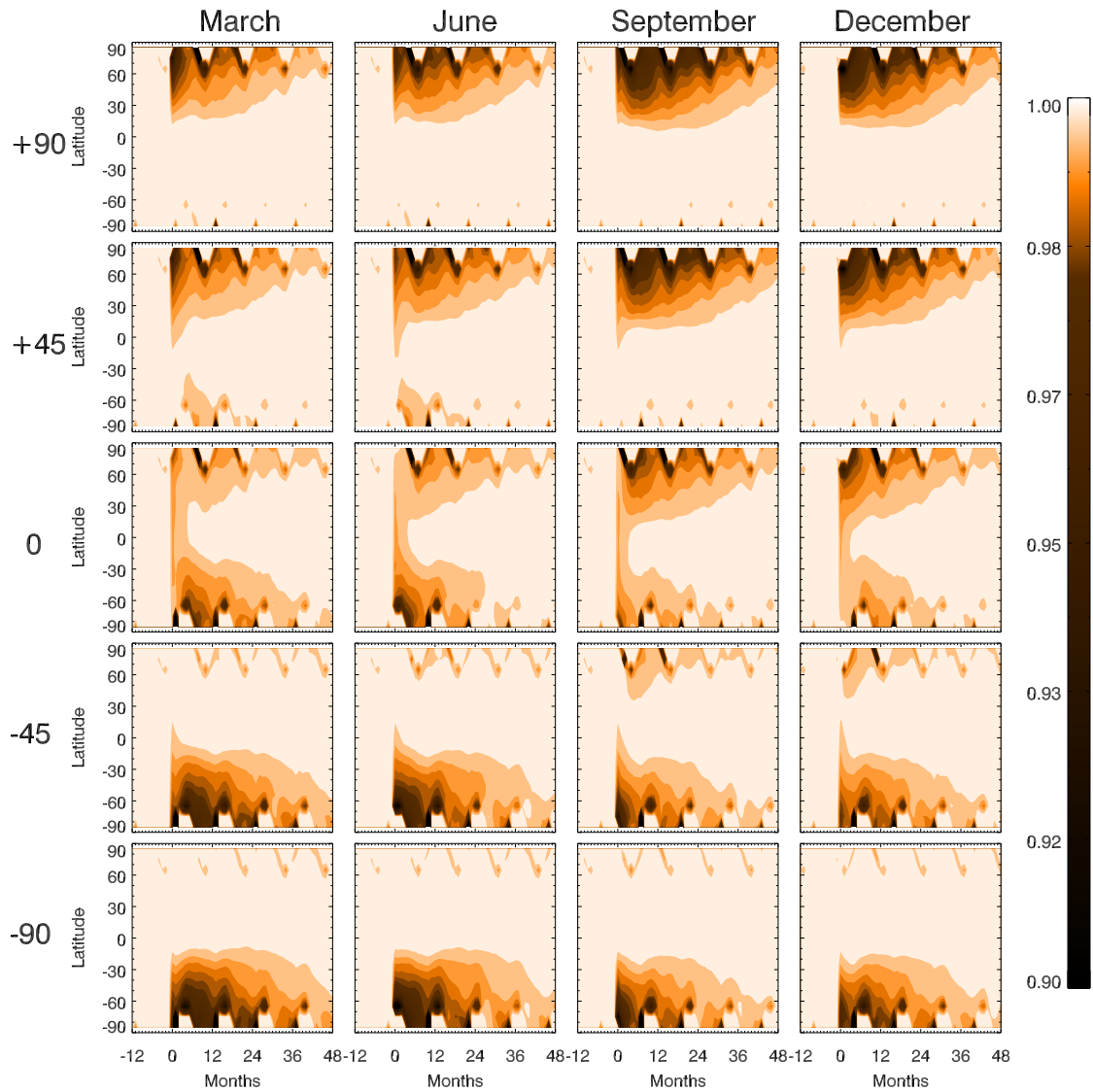}
\caption{The time development of relative solar fluence reaching the surface for $100~\mathrm{kJ ~m^{-2}}$ 
bursts over latitudes $+90^\circ$, $+45^\circ$, the equator, $-45^\circ$, and $-90^\circ$, at the equinoxes and 
solstices.
Relative fluence is the computed amount divided by that in the absence of $\mathrm{NO_2}$ absorption.  
The value is set to 1 for total polar darkness.
\label{fig:opac_grid} }
\end{figure}

An additional expected effect is an increase in atmospheric opacity in the visible due to the accumulation of
 $\mathrm{NO_2}$, a brown gas.  This may lead to climate cooling due to a decrease in surface heating.  We have
 computed the relative solar fluence in the presence of the increased levels of  $\mathrm{NO_2}$ in our model
 runs.  These results are shown in Figure~\ref{fig:opac_grid}.  It is apparent that the greatest decrease in 
solar fluence occurs in the polar regions, corresponding to locally high levels of $\mathrm{NO_2}$.  This may 
be significant for climate cooling since it would lead to a reduction in melting of polar ice during summer.

\begin{figure}
\includegraphics{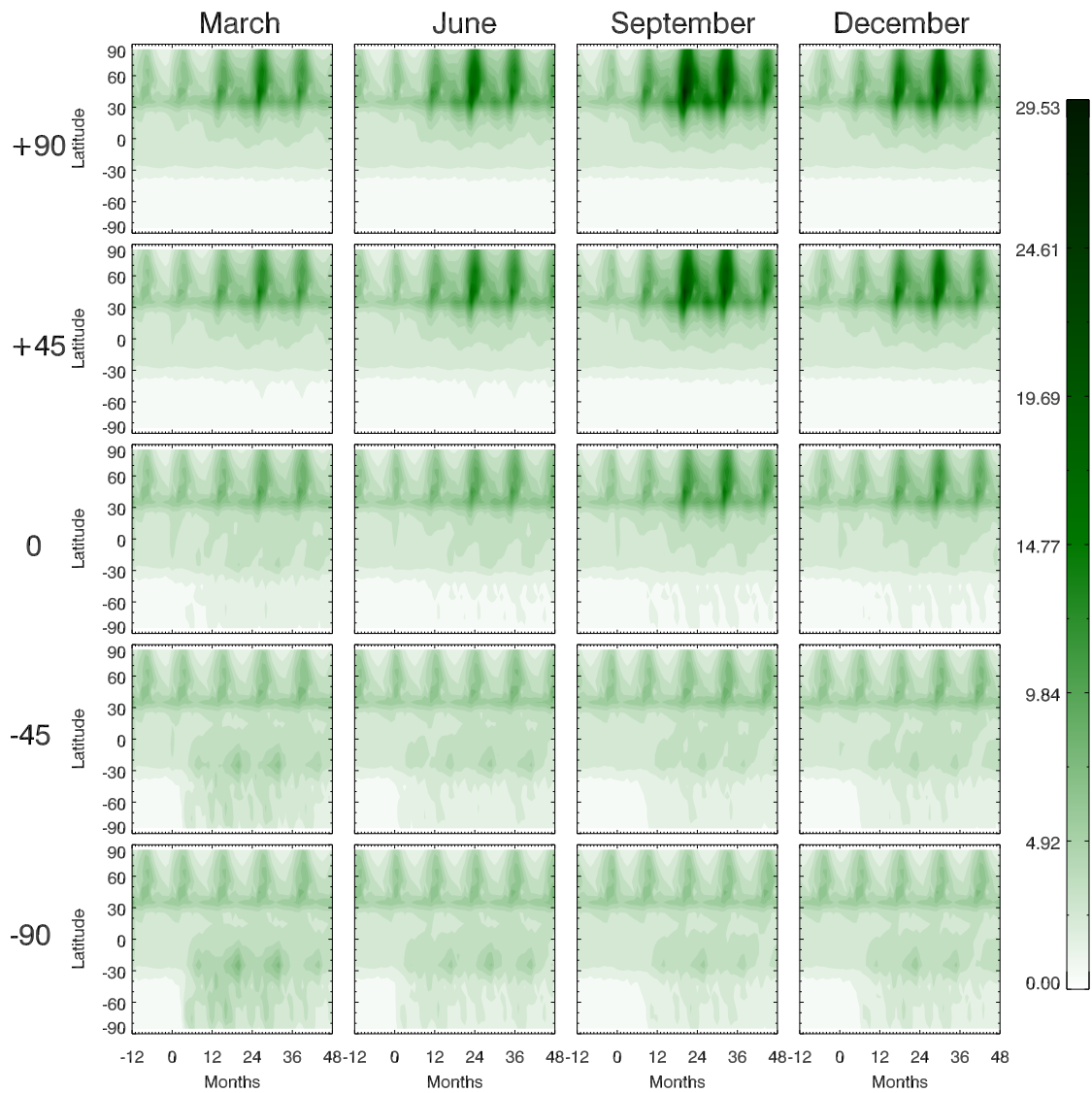}
\caption{Rainout of N as nitrate ($\mathrm{NO_3^-}$) in units of $10^{-10}~\mathrm{g ~m^{-2} ~s^{-1}}$ for 
$100~\mathrm{kJ ~m^{-2}}$ bursts over latitudes $+90^\circ$, $+45^\circ$, the equator, $-45^\circ$, and $-90^\circ$,
 at the equinoxes and solstices.
The pre-burst background includes all non-anthropogenic rainout at modern times, and would be somewhat lower 
prior to the advent of terrestrial nitrogen-fixing plants.
\label{fig:nitrates}}
\end{figure}

Precipitation of nitric acid, $\mathrm{HNO_3}$, is the primary mechanism by which the atmosphere returns to 
normal after a burst.  This precipitation may have complicated effects, as mentioned above, either as an acid 
stress or a fertilizer.  We have computed nitrate deposition using precipitation of $\mathrm{HNO_3}$ as 
determined by the atmospheric model.  Figure~\ref{fig:nitrates} shows nitrate rainout rates (flux of N as
 nitrate) pre- and post-burst for the same cases discussed above.  Much of the latitude asymmetry in this case 
is due to the fact that there is presently more precipitation in the northern hemisphere, especially north of
 about $45^{\circ}$ latitude, as compared to that south of $45^{\circ}$ latitude.  Therefore, these results may
 not be applicable to other eras in the Earth's history when landmasses were in different configurations. 
 Deposition values for some burst cases are similar to or somewhat greater than the total generated ordinarily
 by lightning ($6\times10^{-3}~\mathrm{g~m^{-2}yr^{-1}}$) and other non-biogenic sources \cite{sch97} and values for all 
cases are at least the same order of magnitude as these sources.  Currently, in sensitive ecosystems a flux of $1.5~\mathrm{g~m^{-2}yr^{-1}}$ is considered above the ``critical load'' to cause damage \cite{pho06}.  Smaller inputs may actually be beneficial as fertilizer.  We do not expect a dramatic effect for events in the last 300 My
from nitric acid precipitation, though even small amounts of nitrogen can be of significant fertilizer benefit to the 
biota \cite{sch97}.
Biogenic nitrate was much smaller in the past, so we would expect this effect to be more important then, particularly around the time life began to colonize the land (350-450 Mya).  

In addition to the $100 ~\mathrm{kJ ~m^{-2}}$ received fluence used to produce the above results, we have also 
investigated the impact of different fluence values.  For a fluence of $10 ~\mathrm{kJ ~m^{-2}}$ (corresponding 
to our ``typical'' burst at a distance of about 6.5 kpc) the impact on ozone and subsequent effects is 
significantly reduced.  Maximum localized depletions of around 39\% are seen, as compared to 74\% for the 
higher fluence case.  Globally averaged  depletion reaches about 16\% as compared to 38\% for the higher 
fluence case.  Subsequent DNA damage in the $10 ~\mathrm{kJ ~m^{-2}}$ case is also lower, around 6 at maximum
 as compared to about 15 in the higher fluence case.  This maximum is also seen to have a shorter duration.

While there is significant uncertainty in applying our model to fluences above a few hundred
 $\mathrm{kJ ~m^{-2}}$, we have also investigated a $1 ~\mathrm{MJ ~m^{-2}}$ case as an upper limit.  This
 fluence corresponds to our ``typical'' burst at a distance of about 600 pc.  For this case we find localized 
depletions of ozone up to 80\% and globally averaged depletion up to 65\%.  

We summarize the impact on $\mathrm{O_3}$ for the three fluence cases discussed in Table~\ref{tb2}.  Each case 
is for a burst occurring in late March (equinox) over the Equator.  We point out (as noted in \cite{thom05b}) 
that the globally averaged percent change of ozone scales roughly as the cube root of the fluence.  We also 
note that energy is a more important factor than burst duration in the impact on ozone (see \cite{thom05b} 
and also the next section below).

\section{Continuing Work}
In addition to various fluence values, we have also studied the impact on our results of varying the GRB 
duration and peak energy value, and are currently working to extend the spectral range used.  
Preliminary results \cite{ejz06} indicate that ozone depletion varies by only a few percent with burst duration 
over a wide range of prompt burst times from $\mathrm{10^{-1}~s}$ to $\mathrm{10^8~s}$.  On the other hand,
 harder photon spectra produce greater atmospheric effects from 2 keV up to about 20 MeV, because of greater 
penetration into the stratosphere, with little additional effect for energies higher than 20 MeV.  
This is primarily due to the declining interaction cross section between the photons
and the atmosphere for increasing energy, which allows the photons to penetrate more deeply into the stratosphere,
where the ozone shield resides.  Above MeV scale energies, this cross section becomes nearly constant.
Because of this, a ``saturation'' point is reached as the peak energy approaches 20 MeV, where higher peak 
energy does not result in increased ozone impact.

The overall implication is that total burst fluence and photon energy are more important than burst duration.  
The relative unimportance of duration is primarily due to the long 
timescales of the atmospheric chemistry processes which deplete ozone, compared to the burst durations
 (hours or days vs. minutes), making all these bursts effectively impulsive events.  This fact, combined with
 the greater impact of harder spectra indicates that even less luminous, shorter duration SHB events need to 
be considered along with the long bursts investigated to date, particularly given their likely higher rates at 
low redshift.

\section{Conclusions}
It is clear from results reviewed here, combined with the likelihood of a significant GRB event in the last Gy
 that these events must be considered when looking at the long-term history of life on Earth.  In addition, in 
investigating probable hazards elsewhere in the Galaxy, GRBs must be taken into consideration, along with 
supernovae, impactors and other such events.  As we seek to improve our understanding of where and when life may
 exist beyond the confines of the Earth, these considerations will become increasingly important.

\section{Acknowledments}  We are grateful for research support under NASA grant NNG04GM14G, in the program 
Astrobiology:
Exobiology and Evolutionary Biology. Figures shown here have previously appeared in Astrophysical Journal
 \cite{thom05b} and are used by permission of the author and the publisher.  We thank Bruce Twarog and an anonymous referee for useful comments.


\References
\bibitem[Alvarez et al. (1980)]{alv80} Alvarez, LW et al. 1980 {\it Science} {\bf 208} 1095
\bibitem[Atoyan et al. (2006)]{atoy06} Atoyan, A, Buckley, J, \& Krawczynski, H  2006, {\it Astrophys. J. Lett.}, accepted for publication; astro-ph/0509615
\bibitem[Band et al. (1993)]{band93} Band, D, et al.  1993  {\it Astrophys. J.} {\bf 413} 281
\bibitem[Barthelmy et al. (2005)]{ba05} Barthelmy, SD et al. 2005 {\it Nature} {\bf 438} 994
\bibitem[Cullen et al. (1992)]{cull92} Cullen, JJ, Neale, PJ, \& Lesser, MP  1992
	{\it Science} {\bf 258} 646
\bibitem[Dermer and Holmes (2005)]{dh05} Dermer, CD and Holmes, JM 2005 {\it Astrophys. J. Lett.} {\bf 628} L21
\bibitem[Ejzak et al. (2006)]{ejz06} Ejzak, L, Melott, A, Medvedev, M, \& Thomas, B  2006, in preparation
\bibitem[Freeman et al. (1989)]{free89} Freeman, SE, et al.  1989 {\it Proc. Natl. Acad. Sci.} {\bf 86} 5605
\bibitem[Gehrels et al. (2004)]{geh04} Gehrels, N et al. 2004 {\it Astrophys. J.} {\bf 611} 1005
\bibitem[Gonzalez et al. (2001)]{gonz01} Gonzalez, G, Brownlee, D, \& Ward, P  2001 {\it Icarus} {\bf 152} 185
\bibitem[Guetta \& Piran (2005)]{gp05} Guetta, D, \& Piran, T 2005 {\it Astron. Astrophys.} {\bf 435} 421
\bibitem[Guetta et al. (2005)]{guet05} Guetta, D, Piran, T, \& Waxman, E  2005 {\it Astrophys. J.} {\bf 619} 412 
\bibitem[Hurley et al. (2005)]{hurley05} Hurley, K et al 2005 {\it Nature} {\bf 434} 1098
\bibitem[Ibata et al. (1997)]{iw97} Ibata, RA et al.  1997 {\it Astron. J.} {\bf 113} 634
\bibitem[Jagger (1985)]{jag85} Jagger, J  1985, \emph{Solar-UV Actions on Living Cells} 
	(New York: Praeger) 
\bibitem[Kouwenberg et al. (1999)]{kou99} Kouwenberg, JHM, Browman, HI, Runge, JA, 
	Cullen, JJ, Davis, RF, \& St-Pierre, J-F  1999 {\it Marine Biology} {\bf 134} 285
\bibitem[Langer and Norman (2006)]{ln06} Langer, N, \& Norman, C A 2006 {\it Astrophys. J. Lett.} {\bf 638} L63
\bibitem[Law et al. (2005)]{lj05} Law, DR et al.  2005 {\it Astrophys. J.} {\bf 619} 807
\bibitem[McKinlay \& Diffey (1987)]{mck87} McKinlay, AF \& Diffey, BL  1987, in 
	\textit{Human Exposure to Ultraviolet Radiation: Risks and Regulations}, eds.
\bibitem[Melott et al. (2005)]{mel05} Melott, A, Thomas, B, Hogan, D, 
	Ejzak, L and Jackman, C  2005 {\it Geophys. Res. Lett.} {\bf 32} L14808 doi:10.1029/2005GL02307
\bibitem[Melott et al. (2004)]{mel04} Melott, A, et al.  2004a {\it Int. J. Astrobiology} {\bf 3} 55 (astro-ph/0309415)
\bibitem[Nakar et al. (2005)]{ng05} Nakar, E, Gal-Yam, A, \& Fox, DB  2005 preprint astro-ph/0511254
\bibitem[Phoenix et al. (2006)]{pho06} Phoenix, G, et al.  2006 {\it Global Change Biology} {\bf 12} 470
\bibitem[Preece et al. (2000)]{preece00} Preece, RD, et al.  2000 {\it Astrophys. J. Supp.} {\bf 126} 19
\bibitem[Reid and McAfee (1978)]{rei78} Reid, GC, and  JR McAfee (1978) {\it Nature} {\bf 275}, 489
\bibitem[Setlow (1974)]{set74} Setlow, RB  1974 {\it Pro. Nat. Acad. Sci. USA} {\bf 71} 3363
\bibitem[Setlow et al. (1993)]{set93} Setlow, RB, Grist, E, Thompson, K, 
	\& Woodhead, AD  1993 {\it Proc. Natl. Acad. Sci. US} {\bf 90} 6666
\bibitem[Scalo and Wheeler (2002)]{sw02} Scalo, J and Wheeler, JC 2002 {\it Astrophys. J.} {\bf 566} 723
\bibitem[Schlesinger (1997)]{sch97} Schlesinger, WH 1997 {\it Biogeochemistry}, (2nd Ed.; San Diego:
	Academic Press)
\bibitem[Smith et al. (2004)]{smith04} Smith, DS, Scalo, J, \& Wheeler, JC
	2004 {\it  Icarus} {\bf 171} 229
\bibitem[Smith et al. (1980)]{smith80} Smith, R, Baker, K, Holm-Hansen, O, 
	and Olson, R  1980 {\it Photochem. Photobiol.} {\bf 31} 585
\bibitem[Stanek et al. (2006)]{sg06} Stanek, KZ et al.  2006  preprint astro-ph/0604113
\bibitem[Tanvir et al. (2005)]{tan05} Tanvir, N, et al. 2005 {\it Nature} {\bf 438} 991
\bibitem[Thomas et al. (2005a)]{thom05a} Thomas, B, et al.  2005 {\it Astrophys. J. Lett.} {\bf 622} L153
\bibitem[Thomas et al. (2005b)]{thom05b} Thomas, B, et al. 2005 {\it Astrophys. J.} {\bf 634} 509
\bibitem[Thorsett (1995)]{thor95} Thorsett, S  1995 {\it Astrophys. J. Lett.} {\bf 444} L53
\bibitem[van Loon et al. (2003)]{vl03} van Loon, J, et al.  2003 {\it MNRAS} {\bf 338} 857
\bibitem[Wainwright et al. (2005)]{wb05} Wainwright, C et al. 2005 {\it Astrophys J.}, submitted; astro-ph/0508061
\bibitem[Yong et al. (2006)]{yc05} Yong, D, et al.  2006 {\it Astron. J.} {\bf 131} 2256
\endrefs


\begin{table}
\caption{\label{tb1} Probability of impacting GRB in the last Gy.}
\begin{indented}
\item[] \begin{tabular}{@{}cc}
\br
Distance (kpc) & Probability \\
\mr
1 & 18\% \\
2 & 55\% \\
4 & 70\% \\
\br
\end{tabular}
\end{indented}
\end{table}

\begin{table}
\caption{\label{tb2} Maximum percent change in $\mathrm{O_3}$ for a burst in March over the Equator. }
\begin{indented}
\item[] \begin{tabular}{@{}ccc}
\br
Fluence ($\mathrm{kJ~m^{-2}}$) & Globally Averaged & Localized \\
\mr
10 & -16 & -28 \\
100 & -36 & -55 \\
1000 & -65 & -80 \\
\br
\end{tabular}
\end{indented}
\end{table}

\end{document}